\documentclass[aps,pra,preprint,showpacs,showkeys]{revtex4-1}
\usepackage{amsmath}
\usepackage{amssymb}
\usepackage{graphicx}
\usepackage{bm}

\begin{document}

\title{Maxwell-Sylvester Multipoles and the Geometric Theory of Irreducible Tensor Operators of Quantum Spin Systems}

\author{Patrick Bruno}
\email[]{patrick.bruno@esrf.fr}
\affiliation{European Synchrotron Radiation Facility, BP 220, 38043 Grenoble Cedex, France}


\begin{abstract}
A geometric theory of the irreducible tensor operators of quantum spin systems. It is based upon the Maxwell-Sylvester geometric representation of the multipolar electrostatic potential. In the latter, an order-$\ell$ multipolar potential is represented by a collection of $\ell$ equal length vectors, i.e. by $\ell$ points on a sphere, instead of by its components on some fixed (but arbitrary) basis.  The geometric representation offers a much more appropriate tool for getting physical insight on specific characteristics of a multipole, such as its symmetries, or its departure from ideal symmetry. We derive explicit expressions enabling to perform any calculations we may need to perform on multipoles. All relevant quantities are eventually expressed in terms of scalar products of pairs of vectors (i.e., in terms of geometric quantities such as lengths and angles). The whole formalism is entirely independent of any particular choice of coordinate, and needs no use of the somehow abstract formalism traditionally used when dealing with angular momenta.

The formalism is then applied to treat the problem of the irreducible tensor operators of quantum spin systems. It enables to completely dispense with the calculation and use of the Stevens operators, which can be quite complicated even for moderate values of $\ell$. Explicit expressions for the calculation of expectations values of physical observables are derived. They essentially consist in combinations of scalar products of vector pairs. Together with the coherent state representation of the quantum states of spin systems, this provides a complete geometric, coordinate-free, description of the states, dynamics and physical properties of these systems.
\end{abstract}


\maketitle

\begin{widetext}
\section{Introduction}
For many problems in physics and outside physics, a geometric representation of information is frequently more insightful than a numeric formulation of the same information, no matter how precise, accurate, and otherwise useful the latter may be. To use an analogy, a list of the GPS coordinates of the main cities of a country would hardly provide any insight on its geography and on the opportunities and challenges it implies, whereas a map of the country displaying the position of those cities would instantaneously convey to the observer a global understanding of the geography and its implications.

Since the seminal work of Heisenberg, Born and Jordan \cite{heisenberg1925, born_jordan1925, born_heisenberg_jordan1925}, the algebraic formulation of quantum mechanics has been the major theoretical tool of quantum theory. This algebraic approach has indeed a considerable number of advantages such as its formal elegance and its convenience of use. In particular, due to the availability of very efficient algorithm for performing numerical linear algebra calculations (solving linear equation systems, eigenvalue problems, determinants, etc.) it has been extremely successful for the quantitative numerical study of quantum mechanical systems. On the other hand, a major weakness of the algebraic approach lies in the fact that the physical interpretation of information encoded as complex components of a Hilbert space vector, or as complex matrix elements of an operator is (except in the simplest cases) not immediate at all. For example, if the quantum state of a system exhibits some fundamental symmetry property (such as rotational symmetry, point-group spacial symmetry, time-reversal invariance, etc.) that symmetry may be completely obscured and would require a precise numerical check to be revealed. 

By contrast with the algebraic approach, a geometric formulation of physical laws frequently offers a unique and powerful insight. The most striking example of the power of the geometric approach is Einstein's reformulation and relativistic generalization of the theory of gravitation, in the framework of a pseudo-riemannian geometry of space-time. A further example is the modern geometric formulation of classical mechanics in the language of symplectic geometry on phase space \cite{arnold1989}. Actually, it is quite natural that physical (or more generally scientific) questions be addressed in some geometric setting, since most frequently one has to address such as "how far is the outcome of experiment A from that of experiment B?" of "how far is the outcome of experiments from the prediction of theory?", which naturally calls for the introduction some appropriate metric to quantify to observed "farness". Such questions become most acute when precision is an central issue, such as in metrology. A good survey of the importance of modern geometrical concepts in physics is given in the books of Nakahara and Frankel \cite{nakahara2003,frankel2004}.

In the field of quantum mechanics, geometrical concepts have played an increasingly important r\^{o}le, since Berry's discovery of the geometrical phase \cite{berry1984, aharonov_anandan1987, shapere_wilczek1989, bohm2003, chrucinski2004}.
In particular, the spin coherent states \cite{arecchi1972, gilmore1975, perelomov1986} and the Majorana stellar representation \cite{majorana1932,bloch1945,schwinger1952} for quantum spin systems enable a coordinate-free geometric representation of the pure quantum states of such systems. A more detailed presentation of these geometric description of spin quantum states is given in Sec.~\ref{sec_coherent_majorana}.

The description of operators representing physical observable has remained, so far, almost exclusively based upon their expansion on some basis set, such as the tabulated Stevens operators \cite{rudowicz2004}. The aim of the present paper is to present an alternative approach, in which both the quantum states and the operators acting on them are described in terms of geometrical concepts, and in which calculations of physical quantities can be performed in terms of geometrical quantities such as lengths, area, and angles. 

In Sec.~\ref{sec_tensors}, a coordinate-free tensor formalism is presented. The one-to-one correspondence between symmetric tensors, homogenous polynomials, and psherical functions is stressed, as well as the one-to-one correspondence between traceless symmetric tensors, harmonic polynomials, harmonic potentials, and spherical harmonics, is stressed. The canonical decomposition of symmetric tensors of arbitrary rank into their harmonic (or traceless, or irreducible) components is presented, and explicit formulas for this canonical decomposition are given. Various formulas are derived, enabling to perform calculations within the tensor formalism.

Sec.~\ref{sec_maxwell_sylvester} introduces the Maxwell-Sylvester geometric description of multipolar potentials. In particular, it physical meaning, in terms of the electrostatic interaction among multipoles, which had remained unnoticed so far, is highlighted.

Sec.~\ref{sec_coherent_majorana} presents the geometric description of quantum spin states based upon the spin coherent states, and on the Majorana stellar representation.

Sec.~\ref{sec_geometric_operators} is devoted to the geometric representation of quantum observable operators, in terms of the Maxwell-Sylvester multipoles. Their use for actual calculations of physical quantities is detailed.

Finally, the Appendix details some notations and conventions used in the papers, as well as formulae concerning Legendre polynomials, of which we make use throughout the paper.

\section{Tensors, Homogeneous Polynomials, Spherical Functions, and their Canonical Decomposition}\label{sec_tensors}

\subsection{Symmetric Tensors and Homogeneous Polynomials}


Let $\mathbf{v}_i$ with $i\in [1,n]$ be vectors in $\mathbb{R}^3$. For short-hand notation, we introduce the $n$-tuple $\mathbf{V}^{(n)}$ to describe the ordered collection of the $\mathbf{v}_i$, i.e., $\mathbf{V}^{(n)}\equiv \{\mathbf{v}_1,\mathbf{v}_2, \dots , \mathbf{v}_n \}$. 

A rank-$n$ tensor $\mathcal{T}^{(n)}$ is a (real or complex) function of $\mathbf{V}^{(n)}$ which is linear with respect to each of the $\mathbf{v}_i$'s, i.e., which satisfies
\begin{equation}
\mathcal{T}^{(n)}(\lambda_1\mathbf{v}_1, \dots ,\lambda_n\mathbf{v}_n ) = \lambda_1 \dots \lambda_n \mathcal{T}^{(n)}(\mathbf{v}_1, \dots ,\mathbf{v}_n ).
\end{equation}
A rank-1 tensor is of course just a vector. To make the link with the more familiar description of tensors, which makes use of Cartesian coordinates, let us pick a direct orthonormal triad of (real) unit vectors $(\hat{\mathbf{e}}_1,\hat{\mathbf{e}}_2,\hat{\mathbf{e}}_3)$. The familiar Cartesian components of the tensor $\mathcal{T}^{(n)}$ are then given by
\begin{equation}
\mathcal{T}^{(n)}_{\mu_1, \dots , \mu_n} = \mathcal{T}^{(n)}(\mathbf{e}_{\mu_1},\dots,  \mathbf{e}_{\mu_n})
\end{equation}
with $\mu_i = 1,2,3$. 

For any tensors $\mathcal{A}^{(n)}$ and $\mathcal{B}^{(m)}$, the tensor product $\mathcal{A}^{(n)}\otimes\mathcal{B}^{(m)}$ is the rank-$(n+m)$ tensor defined by 
\begin{equation}
\left( \mathcal{A}^{(n)}\otimes\mathcal{B}^{(m)} \right) (\mathbf{v}_1, \dots ,\mathbf{v}_n,\mathbf{v}_{n+1}, \dots , \mathbf{v}_{n+m} ) \equiv \mathcal{A}^{(n)}( \mathbf{v}_1, \dots ,\mathbf{v}_n ) \mathcal{B}^{(m)} ( \mathbf{v}_{n+1}, \dots ,\mathbf{v}_{n+m} ) .
\end{equation}
\begin{equation}
\left( \mathcal{A}^{(n)}\otimes\mathcal{B}^{(m)} \right) (\mathbf{V}^{(n)},\mathbf{W}^{(m)}) \equiv \mathcal{A}^{(n)}(\mathbf{V}^{(n)}) \mathcal{B}^{(m)} (\mathbf{W}^{(m)}) .
\end{equation}
Let $S_n$ be the symmetric group of index $n$; 
for any permutation $\sigma \in S_n$, we define $\sigma\mathbf{V}^{(n)} \equiv \{\mathbf{v}_{\sigma(1)},\mathbf{v}_{\sigma(2)}, \dots , \mathbf{v}_{\sigma(n)} \}$. A tensor $\mathcal{T}^{(n)}$ is said to be fully symmetric if for any permutation $\sigma \in S_n$, it satisfies
\begin{equation}
\mathcal{T}^{(n)}(\sigma\mathbf{V}^{(n)}) = \mathcal{T}^{(n)}(\mathbf{V}^{(n)}) .
\end{equation}
To any tensor $\mathcal{T}^{(n)}$, we can associate a fully symmetric tensor $\text{Sym}\underline{\mathcal{T}}^{(n)}$, defined by
\begin{equation}
\text{Sym}\mathcal{T}^{(n)}(\mathbf{V}^{(n)}) =\frac{1}{n!} \sum_{\sigma \in S_n} \mathcal{T}^{(n)}(\sigma\mathbf{V}^{(n)}) .
\end{equation}
Let us introduce the fully symmetrized product $\mathcal{A}^{(n)}\bigodot\mathcal{B}^{(m)}$, defined by
\begin{equation}\label{sym product}
\mathcal{A}^{(n)}\bigodot\mathcal{B}^{(m)} \equiv \text{Sym}\left( \mathcal{A}^{(n)}\otimes\mathcal{B}^{(m)}\right) ,
\end{equation}
and the tensor power 
\begin{equation}
\begin{array}{c}
\left( \mathcal{A}^{(n)}\right)^{\odot k} \equiv \underbrace{\mathcal{A}^{(n)}\odot \dots \odot \mathcal{A}^{(n)} }
\\ \ \ \ \ \ \ \ \ \ \ \ \ \ k \text{ factors},
\end{array}
\end{equation}
which is a tensor of rank $nk$. A tensor of central importance is the rank-2 tensor $\boldsymbol{\delta}^{(2)}$ defined by
\begin{equation}
\boldsymbol{\delta}^{(2)}(\mathbf{v}_1,\mathbf{v}_2) = \mathbf{v}_1\cdot\mathbf{v}_2 .
\end{equation}
Expressed in Cartesian coordinates, it is given by the Kronecker symbol. It is straightforward to prove that it can also be expressed as
\begin{equation}
\boldsymbol{\delta}^{(2)} = \frac{3}{4\pi}\int_{S^2}\!\!\!\mathrm{d}^2 \hat{\mathbf{n}}\ \hat{\mathbf{n}}\odot \hat{\mathbf{n}},
\end{equation}
where the integral runs over the 2-dimensional sphere of unit radius $S^2$. For any rank-$n$ tensor $\mathcal{A}^{(n)}$, we define the rank-$(n-1)$ derived tensor $\mathcal{A}^{(n)} [\underset{i}{\mathbf{v}}]$ obtained by fixing the $i-$th entry to $\mathbf{v}$  and leaving the $n-1$ other entries free. Similarly, we can obtain a rank-$(n-p)$  derived tensor by fixing $p$ entries of a rank-$n$ tensor. The contraction over the pair of indices $(i,j)$ of two tensors $\mathcal{A}^{(n)}$ and $\mathcal{B}^{(m)}$ can thus be defined as
\begin{equation}
\mathcal{A}^{(n)} \underset{(i,j)}{\,\colon} \mathcal{B}^{(m)} \equiv \frac{3}{4\pi}\int_{S^2}\!\!\!\mathrm{d}^2 \hat{\mathbf{n}}\ \mathcal{A}^{(n)} [\underset{i}{\hat{\mathbf{n}}} ] \otimes \mathcal{B}^{(m)}[\underset{j}{\hat{\mathbf{n}}}] . 
\end{equation}
This is a rank-$(n+m-2)$ tensor. This process can be iterated to define the $p$-fold contraction over $p$ pairs of indices. It is easy to get convinced that the contraction defined in this way coincides with familiar definition in terms in terms of summation over Cartesian indices.

Clearly, when dealing with fully symmetric tensors, the choice of the entries when defining derived tensors and contractions is unimportant, and we shall simplify the notations by omitting the indices $i,j,\dots$. So for the $p$-fold contraction of two symmetrical tensors will be written
\begin{equation}
\mathcal{A}^{(n)} \underset{(p)}{\,\colon} \mathcal{B}^{(m)} \equiv \left( \frac{3}{4\pi}\right)^p \prod_{k=1}^{p}\left( \int_{S^2}\!\!\!\mathrm{d}^2 \hat{\mathbf{n}}_k\right) \ \mathcal{A}^{(n)} [\hat{\mathbf{n}}_1, \dots, \hat{\mathbf{n}}_p ] \otimes \mathcal{B}^{(m)} [\hat{\mathbf{n}}_1, \dots, \hat{\mathbf{n}}_p ]. 
\end{equation}
It is a tensor of rank $(n+m-2p)$. The 1-fold contraction of two symmetric tensors is the inner product:
\begin{equation}
\mathcal{A}^{(n)} \cdot \mathcal{B}^{(m)} \equiv \mathcal{A}^{(n)} \underset{(1)}{\,\colon} \mathcal{B}^{(m)} ,
\end{equation}
and one easily check that $\boldsymbol{\delta}^{(2)}$ is the unit for the inner product, i.e.,
\begin{equation}
\mathcal{A}^{(n)} \cdot \boldsymbol{\delta}^{(2)} = \boldsymbol{\delta}^{(2)}\cdot \mathcal{A}^{(n)} =\mathcal{A}^{(n)} .
\end{equation}
The notation of the full contraction of two symmetrical tensors (i.e., for $p = \text{min}(n,m)$), will be further simplified and written $\mathcal{A}^{(n)} \colon \mathcal{B}^{(m)}$, hich is a tensor of rank $|n-m|$. From the definition of the contraction, one can easily observe that, given three symmetric tensors $\mathcal{A}^{(n)}$, $\mathcal{B}^{(m)}$, $\mathcal{C}^{(n+m)}$, one has the following property:
\begin{eqnarray}\label{dot colon}
\left( \mathcal{A}^{(n)} \odot \mathcal{B}^{(m)} \right) \colon \mathcal{C}^{(n+m)} &=& 
\mathcal{A}^{(n)} \colon \mathcal{D}^{(n)}  \\
&=& \mathcal{B}^{(m)} \colon \mathcal{E}^{(m)} ,
\end{eqnarray}
where
\begin{eqnarray}
\mathcal{D}^{(n)} &\equiv & \mathcal{B}^{(m)} \colon \mathcal{C}^{(n+m)} \\
\mathcal{E}^{(m)} &\equiv & \mathcal{A}^{(n)} \colon \mathcal{C}^{(n+m)} .
\end{eqnarray}

The $(i,j)$-trace of a tensor $\mathcal{A}^{(n)}$ is the rank-$(n-2)$ tensor defined as
\begin{equation}
\text{Tr}_{(i,j)}\mathcal{A}^{(n)}\equiv \frac{3}{4\pi}\int_{S^2}\!\!\!\mathrm{d}^2 \mathbf{n}\ \mathcal{A}^{(n)} [\underset{i}{\hat{\mathbf{n}}},\underset{j}{\hat{\mathbf{n}}} ] .
\end{equation}
For fully symmetrical tensors, we can omit to specify the indices $(i,j)$, and simply write
\begin{equation}
\mathcal{A}^{(n,1)}\equiv \text{Tr}\mathcal{A}^{(n)}\equiv  \mathcal{A}^{(n)} \colon \boldsymbol{\delta}^{(2)} .
\end{equation}
The $p$-fold trace is then
\begin{equation}
\mathcal{A}^{(n,p)}\equiv \mathcal{A}^{(n,p-1)} \colon \boldsymbol{\delta}^{(2)} =  \mathcal{A}^{(n)} \colon (\boldsymbol{\delta}^{(2)})^{\odot p} .
\end{equation}
It is a tensor of rank $(n-2p)$.

A tensor is said to be traceless if it satisfies 
\begin{equation}
\text{Tr}_{(i,j)}\mathcal{A}^{(n)} =0
\end{equation}
for any pair of indices $(i,j)$.

Let us pick some cartesian coordinates $(x,y,z)$ in $\mathbb{R}^3$. A function $f(\mathbf{r})$ on $\mathbb{R}^3$ is a homogeneous polynomial of degree $n$ of the 3 variables $x,y,z$ if it can be expressed as
\begin{equation}
f(\mathbf{r}) = \sum_{p,q,s \ge 0} f_{p,q,r}x^p y^q z^s ,
\end{equation}
with $p+q+s=n$. For any tensor $\mathcal{A}^{(n)}$, we can define a function on $\mathbb{R}^3$ via
\begin{equation}
\mathcal{A}^{(n)}(\mathbf{r})\equiv \mathcal{A}^{(n)}(\mathbf{r}^{\odot n}) = \mathcal{A}^{(n)} \colon (\mathbf{r}^{\odot n}) .
\end{equation}
It is easy to see that it is a homogeneous polynomial of degree $n$, and there is actually a one-to-one correspondence between rank-$n$ fully symmetrical tensors and order-$n$ homogeneous polynomials \cite{Backus1970}. An the correspondence is given by
\begin{equation}
\mathcal{A}^{(n)}(\mathbf{V}^{(n)})= \frac{1}{n!}\prod_{k=1}^{n}\left(  \mathbf{v}_k \cdot \bm{\partial}_\mathbf{r}\right)   \mathcal{A}^{(n)} (\mathbf{r}) .
\end{equation}
From now on, unless explicitly specified, we shall deal only with fully symmetrical tensors, and thus shall identify them with their corresponding homogeneous polynomial. So without any ambiguity, we shall use the same symbol to represent the tensor and the polynomial. The context should make clear which one is actually meant.

Let us now express the $p$-fold trace $\mathcal{A}^{(n,p)}$ of a symmetric tensor in terms of the action of the Laplace operator $\Delta_\mathbf{r}$ on $\mathcal{A}^{(n)}$. From the definition of the trace and the above equation, one has
\begin{eqnarray}
\mathcal{A}^{(n,1)}(\mathbf{V}^{(n-2)})&=& \frac{1}{n!}\prod_{k=1}^{n-2} (\bm{\partial}_\mathbf{r}\cdot \bm{\partial}_\mathbf{r}) \left(  \mathbf{v}_k \cdot \bm{\partial}_\mathbf{r}\right)   \mathcal{A}^{(n)} (\mathbf{r}) \\
&=& \frac{1}{(n-2)!}\prod_{k=1}^{n-2}  \left(  \mathbf{v}_k \cdot \bm{\partial}_\mathbf{r}\right)  \left( \frac{(n-2)!}{n!}\Delta_\mathbf{r} \mathcal{A}^{(n)} (\mathbf{r})\right) , 
\end{eqnarray}
so that
\begin{equation}
\mathcal{A}^{(n,1)}(r) = \frac{(n-2)!}{n!}\Delta_\mathbf{r} \mathcal{A}^{(n)} (\mathbf{r}) .
\end{equation}
By iteration, we finally get
\begin{equation}\label{trace laplacian}
\mathcal{A}^{(n,p)}(r) = \frac{(n-2p)!}{n!} \left( \Delta_\mathbf{r}\right)^p  \mathcal{A}^{(n)} (\mathbf{r}) .
\end{equation}

\subsection{Traceless Symmetric Tensors and Harmonic Polynomials}

A function $f(\mathbf{r})$ over $\mathbb{R}^3$ is said to be harmonic if it satisfies Laplace's equation
\begin{equation}
\Delta_\mathbf{r} f(\mathbf{r}) \equiv \bm{\partial}_\mathbf{r}\cdot \bm{\partial}_\mathbf{r} f(\mathbf{r}) =0 .
\end{equation}
For an order-$n$ polynomial $\mathcal{P}^{(n)} (\mathbf{r})$ associated to the symmetric tensor $\mathcal{P}^{(n)}$, one can easily show that 
\begin{equation}
\Delta_\mathbf{r} \mathcal{P}^{(n)} (\mathbf{r}) = n(n-1)\,\mathcal{P}^{(n,1)} (\mathbf{r}) ,
\end{equation}
so that a polynomial is harmonic iff its associated tensor is traceless. By iteration of the above formula, we get
\begin{equation}
(\Delta_\mathbf{r})^p \mathcal{P}^{(n)} (\mathbf{r}) = \frac{n!}{(n-2p)!}\,\mathcal{P}^{(n,p)} (\mathbf{r}) ,
\end{equation}
Harmonic polynomials are called regular solid harmonics. By Kelvin's transformation theorem \cite{Helms1969}, to each regular solid harmonic $\mathcal{H}^{(\ell )}(\mathbf{r})$, we can associate another solution of Laplace's equation, the irregular solid harmonic defined by
\begin{equation}
\mathcal{V}^{(\ell )}(\mathbf{r}) \equiv \frac{1}{r}\mathcal{H}^{(\ell )}\left( \frac{\mathbf{r}}{r^2}\right) = \frac{1}{r^{2\ell +1}} \mathcal{H}^{(\ell )}(\mathbf{r}) .
\end{equation}
As we shall discuss later, $\mathcal{V}^{(\ell )}(\mathbf{r})$ can be thought of as the electrostatic potential of an electric $2^\ell$-pole located at $\mathbf{r}=0$. The spherical harmonics are defined as
\begin{equation}\label{potential harmonic}
\mathcal{Y}^{(\ell )}(\mathbf{r}) \equiv \frac{\mathcal{H}^{(\ell )}(\mathbf{r})}{r^\ell} = r^{2\ell +1} \mathcal{V}^{(\ell )}(\mathbf{r}) .
\end{equation}
The order-$\ell$ spherical harmonics satisfy 
\begin{equation}
\Delta_\mathbf{r}\mathcal{Y}^{(\ell )}(\mathbf{r}) + \frac{\ell (\ell +1)}{r^2}\mathcal{Y}^{(\ell )}(\mathbf{r}) = 0 .
\end{equation}
Writing $\mathcal{H}^{(\ell )}(\mathbf{r}) = r^\ell \mathcal{Y}^{(\ell )}(\mathbf{r})$ and using the fact 
that $\mathbf{r}\cdot \bm{\partial}_\mathbf{r} \mathcal{Y}^{(\ell )}(\mathbf{r}) =0$, one can show easily that the solid harmonics satisfy
\begin{equation}
\Delta_\mathbf{r}\left(  (\mathbf{r}\cdot \mathbf{r})^n \mathcal{H}^{(\ell )}(\mathbf{r}) \right) 
= 2n (2n+2\ell+1) (\mathbf{r}\cdot \mathbf{r})^{n-1} \mathcal{H}^{(\ell )}(\mathbf{r}) .
\end{equation}
By iteration, we get
\begin{eqnarray}\label{laplacian}
\left( \Delta_\mathbf{r}\right)^p \left(  (\mathbf{r}\cdot \mathbf{r})^n \mathcal{H}^{(\ell )}(\mathbf{r}) \right) 
&=& \frac{(2n)!!}{(2(n-p))!!}\, \frac{(2(n+\ell)+1)!!}{(2(n+\ell -p)+1)!!} (\mathbf{r}\cdot \mathbf{r})^{n-p} \mathcal{H}^{(\ell )}(\mathbf{r})\ \text{for }\ \ p\le n , \\
&=& 0\ \text{for }\ \ p >n .
\end{eqnarray}
In view of the one-to-one correspondence between traceless symmetric tensors, regular and irregular solid harmonics, and spherical harmonics, we shall use them interchangeably, depending on which is more appropriate in the given context.

We define later use, we introduce the inner product of two spherical functions $f(\hat{\mathbf{n}})$ and $g(\hat{\mathbf{n}})$
\begin{equation}
\left\langle f | g\right\rangle \equiv \frac{1}{4\pi} \int_{S^2}\!\!\!\mathrm{d}^2 \hat{\mathbf{n}}\ f^\star(\hat{\mathbf{n}})\ g(\hat{\mathbf{n}}) .
\end{equation}

\subsection{Canonical decomposition of symmetric tensors}\label{section canonical}

Given a symmetric rank-$n$ tensor $\mathcal{A}^{(n )}$, we seek a decomposition into its harmonic (traceless) components as:
\begin{equation} \label{canonical decomposition}
\mathcal{A}^{(n )} = \sum_{k =0}^{\lfloor n/2 \rfloor} 
\mathcal{A}^{(n)}_{(n-2k)} \odot (\boldsymbol{\delta}^{(2)})^{\odot k} .
\end{equation}
In the above expression, the rank-$(n-2k)$ tensor $\mathcal{A}^{(n)}_{(n-2k)}$ is the rank-$(n-2k)$ harmonic (traceless) component of $\mathcal{A}^{(n )}$. The existence and unicity of such decomposition will be proven below. 

Rewriting (\ref{canonical decomposition}) as
\begin{equation} \label{canonical decomposition2}
\mathcal{A}^{(n )} (\mathbf{r})= \sum_{k =0}^{\lfloor n/2 \rfloor} (\mathbf{r}\cdot \mathbf{r})^k
\mathcal{A}^{(n)}_{(n-2k)} (\mathbf{r}) .
\end{equation}
Comparing with the harmonic decomposition of a spherical function, Eqs.~(\ref{spherical harmonic component},\ref{spherical harmonic decomposition}), we get
\begin{equation}
\mathcal{A}^{(n)}_{(n-2k)} (\mathbf{r}) = r^{n-2k} \frac{2(n-2k)+1}{4\pi} \int_{S^2}\!\!\!\mathrm{d}^2 \hat{\mathbf{n}}\
\mathcal{A}^{(n)} (\hat{\mathbf{n}}) P_{n-2k} \left( \hat{\mathbf{n}}\cdot \frac{\mathbf{r}}{r}\right) .
\end{equation}
We now want to derive an explicit expression of $\mathcal{A}^{(n)}_{(n-2k)}$ in terms of the $p$-fold traces of $\mathcal{A}^{(n)}$.  Using Eqs.~(\ref{trace laplacian},\ref{laplacian}), and taking the $p$-fold trace of Eq.~(\ref{canonical decomposition2}), we get
\begin{equation}
\mathcal{A}^{(n,p )} (\mathbf{r})= \sum_{k =p}^{\lfloor n/2 \rfloor}\,
\frac{(2k)!!}{(2(k-p))!!} \, \frac{(2(n-k)+1)!!}{(2(n-k-p)+1)!!}
 (\mathbf{r}\cdot \mathbf{r})^{k-p}
\mathcal{A}^{(n)}_{(n-2k)} (\mathbf{r}) ,
\end{equation}
i.e.,
\begin{equation}\label{linear system}
(\boldsymbol{\delta}^{(2)})^{\odot (p)} \odot\mathcal{A}^{(n,p )}= \sum_{k =p}^{\lfloor n/2 \rfloor}\,
\frac{(2k)!!}{(2(k-p))!!} \, \frac{(2(n-k)+1)!!}{(2(n-k-p)+1)!!}
(\boldsymbol{\delta}^{(2)})^{\odot (k)} \odot
\mathcal{A}^{(n)}_{(n-2k)} .
\end{equation}
We have thus obtained a set of linear equation relating the $p$-fold traces to the harmonic components $\mathcal{A}^{(n)}_{(n-2k)}$ of the canonical decomposition. The $(\lfloor n/2 \rfloor +1 )\times (\lfloor n/2 \rfloor +1 )$ matrix of the linear system is an upper triangular matrix, with strictly positive diagonal elements; it is therefore invertible, which proves the existence and unicity of the canonical decomposition. 

To obtain the explicit expression of the harmonic components, one would need to invert matrix of the linear system. Instead, I shall use here a simpler approach. Let us specialize to the case of of a symmetric tensor given by
\begin{equation}\label{tensor power}
\mathcal{B}^{(n)} \equiv \mathbf{b}^{\odot n} .
\end{equation}
The $p$-fold traces are given by 
\begin{equation}
\mathcal{B}^{(n,p)} = (\mathbf{a}\cdot \mathbf{b})^p\,\mathbf{b}^{\odot (n-p)} .
\end{equation}
We have 
\begin{eqnarray}
\mathcal{B}^{(n)}(\mathbf{r}) &=& (\mathbf{a}\cdot \mathbf{r})^n \\
&=& (br)^n \sum_{k =0}^{\lfloor n/2 \rfloor}\, q_{n,k} P_{n-2k}\left( \frac{\mathbf{b}\cdot \mathbf{r}}{br} \right) ,
\end{eqnarray}
where we have used the expansion of the monomial in Legendre polynomials Eqs.(\ref{monomial},\ref{qnk}). Thus, we have
\begin{equation}
\mathcal{B}^{(n)}_{(n-2k)} (\mathbf{r}) =  q_{n,k} (\mathbf{b}\cdot \mathbf{b})^k \, (br)^{n-2k} \,   P_{n-2k}\left( \frac{\mathbf{b}\cdot \mathbf{r}}{br} \right) .
\end{equation}
It then simply remains to insert in the above equation the expression of the Legendre polynomials (\ref{legendre expansion}), and we finally get
\begin{equation}\label{harmonic decomposition tensor}
\mathcal{B}^{(n)}_{(n-2k)}= q_{n,k}\, \sum_{p =k}^{\lfloor n/2 \rfloor}\, p_{n-2k, p-k} 
(\boldsymbol{\delta}^{(2)})^{\odot (p-k)} \odot
\mathcal{B}^{(n,p)} ,
\end{equation}
which the announced explicit expression of the harmonic components of a symmetric tensor in terms of its traces. Although, we have used the specific case Eq.~(\ref{tensor power}) to derive the above result, it follows from the reasoning leading to Eq.~(\ref{linear system}) that it actually holds for any symmetric tensor.

Let us now use the above result, Eq.~(\ref{harmonic decomposition tensor}) to derive some results we shall need later. Let us consider the symmetric tensor $\mathcal{A}^{(n)} $ given as a symmetrized tensor product of $n$ vectors $\mathbf{a}_i$, i.e.,
\begin{equation}
\mathcal{A}^{(n)} \equiv \bigodot_{i=1}^n \, \mathbf{a}_i .
\end{equation}
We want to calculate the integral 
\begin{equation}
\frac{1}{4\pi} \int_{S^2}\!\!\!\mathrm{d}^2 \hat{\mathbf{n}}\ \prod_{i=1}^{n} (\mathbf{a}_i \cdot \hat{\mathbf{n}})   =
\frac{1}{4\pi} \int_{S^2}\!\!\!\mathrm{d}^2 \hat{\mathbf{n}}\ \mathcal{A}^{(n)}(\hat{\mathbf{n}}).
\end{equation}
Obviously, by symmetry, this is zero if $n$ is odd, so we take $n\equiv 2m$. Using the canonical decomposition (\ref{harmonic decomposition tensor}) and the fact that integral over the sphere of an order-$\ell$ spherical harmonic vanishes if $\ell \neq 0$, we get
\begin{eqnarray}
\frac{1}{4\pi} \int_{S^2}\!\!\!\mathrm{d}^2 \hat{\mathbf{n}}\ \prod_{i=1}^{2m} (\mathbf{a}_i \cdot \hat{\mathbf{n}})   &=& \mathcal{A}^{(2m)}_{(0)}  \\
&=& q_{2m,m}\, p_{0,0} \,\mathcal{A}^{(2m,m)}
\end{eqnarray}
So we finally get 
\begin{equation}
\frac{1}{4\pi} \int_{S^2}\!\!\!\mathrm{d}^2 \hat{\mathbf{n}}\ \prod_{i=1}^{n} (\mathbf{a}_i \cdot \hat{\mathbf{n}})   = \frac{1}{(2m+1)}\, \frac{1}{(2m)!}\sum_{\sigma \in S_{2m}} \prod_{i=1}^m (\mathbf{a}_{\sigma (i)} \cdot \mathbf{a}_{\sigma (i+m)}) .
\end{equation}
There are actually $(2m-1)!!$ different terms in the above sum, corresponding to the number of different  pairings of the $2m$ vectors $\mathbf{a}_i$, so the above result can also be expressed as
\begin{equation}\label{integral pairings}
\frac{1}{4\pi} \int_{S^2}\!\!\!\mathrm{d}^2 \hat{\mathbf{n}}\ \prod_{i=1}^{n} (\mathbf{a}_i \cdot \hat{\mathbf{n}})   = \frac{1}{(2m+1)}\, \frac{1}{(2m-1)!!}\sum_{\text{pairings}}\ \   \prod_{\text{pairs}\ (i,j)} (\mathbf{a}_{i} \cdot \mathbf{a}_{j}) .
\end{equation}
As a byproduct, this yields the following useful formula
\begin{equation}
\frac{1}{4\pi} \int_{S^2}\!\!\!\mathrm{d}^2 \hat{\mathbf{n}}\ {\hat{n}_x}^{2p} \, {\hat{n}_y}^{2q}\, {\hat{n}_z}^{2r} = 
\frac{(2p)!}{p!}\,\frac{(2q)!}{q!}\,\frac{(2r)!}{r!}\, \frac{(p+q+r)!}{(2(p+q+r)+1)!} .
\end{equation}
Let us now consider two rank-$n$ symmetric tensors $\mathcal{A}^{(n)}$ and $\mathcal{B}^{(n)}$, and their respective order-$n$ harmonic components  $\mathcal{A}^{(n)}_{(n)}$ and $\mathcal{B}^{(n)}_{(n)}$. We want to calculate
\begin{equation}
\left\langle \mathcal{A}^{(n)}_{(n)} \,  \mathcal{B}^{(n)}_{(n)} \right\rangle_{S^2} \equiv 
\frac{1}{4\pi} \int_{S^2}\!\!\!\mathrm{d}^2 \hat{\mathbf{n}}\ 
\mathcal{A}^{(n)}_{(n)}(\hat{\mathbf{n}}) \,  \mathcal{B}^{(n)}_{(n)}(\hat{\mathbf{n}}) .
\end{equation}
Using the Cartesian form of the tensors $\mathcal{A}^{(n)}_{(n)}$ and $\mathcal{B}^{(n)}_{(n)}$, and using Einstein's convention of summation over repeated indices, we have
\begin{equation}
\left\langle \mathcal{A}^{(n)}_{(n)} \,  \mathcal{B}^{(n)}_{(n)} \right\rangle_{S^2} \equiv 
{\mathcal{A}^{(n)}_{(n)}}_{\mu_1, \dots, \mu_n} \, {\mathcal{B}^{(n)}_{(n)}}_{\nu_1, \dots, \nu_n} \frac{1}{4\pi} \int_{S^2}\!\!\!\mathrm{d}^2 \hat{\mathbf{n}}\ (\hat{n}_{\mu_1} \dots \hat{n}_{\mu_n})\,
(\hat{n}_{\nu_1} \dots \hat{n}_{\nu_n}) .
\end{equation}
We can calculate the integral using Eq.~(\ref{integral pairings}). We remark that, among the $(2n-1)!!$ pairings involved in the summation, all the pairings in which two $\mu$ indices are paired give a zero contribution, because $\mathcal{A}^{(n)}_{(n)}$ and $\mathcal{B}^{(n)}_{(n)}$ are traceless. So we need to consider only the pairings in which each $\mu$ index is paired with a $\nu$ index. There are $n!$ such pairings, and each over yields a contribution equal to $\mathcal{A}^{(n)}_{(n)}\colon  \mathcal{B}^{(n)}_{(n)}$, so that we finally get
\begin{equation}\label{inner product harmonics}
\left\langle \mathcal{A}^{(n)}_{(n)} \,  \mathcal{B}^{(n)}_{(n)} \right\rangle_{S^2} = \frac{n!}{(2n+1)!!}
\ \mathcal{A}^{(n)}_{(n)}\colon  \mathcal{B}^{(n)}_{(n)} .
\end{equation}

Using the canonical decomposition (\ref{canonical decomposition2}), we have
\begin{equation}
\left\langle \mathcal{A}^{(n)}_{(n)} \,  \mathcal{B}^{(n)}_{(n)} \right\rangle_{S^2} = \frac{(2n+1)^2}{(4\pi)^3} \int_{S^2}\!\!\!\mathrm{d}^2 \hat{\mathbf{n}}\int_{S^2}\!\!\!\mathrm{d}^2 \hat{\mathbf{a}}\int_{S^2}\!\!\!\mathrm{d}^2 \hat{\mathbf{b}}\ 
\mathcal{A}^{(n)}(\hat{\mathbf{a}}) \, 
P_n(\hat{\mathbf{a}}\cdot \hat{\mathbf{n}}) \, 
P_n(\hat{\mathbf{n}}\cdot \hat{\mathbf{b}}) \, 
\mathcal{B}^{(n)}(\hat{\mathbf{b}}) .
\end{equation}
Using the reproducing kernel property of Legendre polynomials (\ref{reproducing kernel}), this becomes
\begin{equation}
\left\langle \mathcal{A}^{(n)}_{(n)} \,  \mathcal{B}^{(n)}_{(n)} \right\rangle_{S^2} = \frac{2n+1}{(4\pi)^2} \int_{S^2}\!\!\!\mathrm{d}^2 \hat{\mathbf{a}}\int_{S^2}\!\!\!\mathrm{d}^2 \hat{\mathbf{b}}\ 
\mathcal{A}^{(n)}(\hat{\mathbf{a}}) \, 
P_n(\hat{\mathbf{a}}\cdot \hat{\mathbf{b}}) \, 
\mathcal{B}^{(n)}(\hat{\mathbf{b}}) ,
\end{equation}
i.e.,
\begin{equation}
\left\langle \mathcal{A}^{(n)}_{(n)} \,  \mathcal{B}^{(n)}_{(n)} \right\rangle_{S^2} = 
\left\langle \mathcal{A}^{(n)}_{(n)} \,  \mathcal{B}^{(n)} \right\rangle_{S^2} =
\left\langle \mathcal{A}^{(n)} \,  \mathcal{B}^{(n)}_{(n)} \right\rangle_{S^2} .
\end{equation}
On the other hand, using the canonical decomposition (\ref{canonical decomposition}), we have
\begin{equation}
\mathcal{A}^{(n)} \colon \mathcal{B}^{(n)}_{(n)} = \sum_{k =0}^{\lfloor n/2 \rfloor} 
\left( \mathcal{A}^{(n)}_{(n-2k)} \odot (\boldsymbol{\delta}^{(2)})^{\odot k} \right) \colon  \mathcal{B}^{(n)}_{(n)} .
\end{equation}
Using the property (\ref{dot colon}) of the tensor contraction, we have 
\begin{equation}
\left( \mathcal{A}^{(n)}_{(n-2k)} \odot (\boldsymbol{\delta}^{(2)})^{\odot k} \right) \colon  \mathcal{B}^{(n)}_{(n)} =  \mathcal{A}^{(n)}_{(n-2k)} \colon \left( (\boldsymbol{\delta}^{(2)})^{\odot k} \colon \mathcal{B}^{(n)}_{(n)} \right)  = 0 \ \text{for} \ k\neq 0,
\end{equation}
because $(\boldsymbol{\delta}^{(2)})^{\odot k} \colon \mathcal{B}^{(n)}_{(n)}$ is the $k$-fold trace of the harmonic (i.e., traceless) tensor $\mathcal{B}^{(n)}_{(n)}$ and therefore vanishes. Thus we have
\begin{equation}
\mathcal{A}^{(n)}_{(n)}\colon  \mathcal{B}^{(n)}_{(n)} = \mathcal{A}^{(n)}\colon  \mathcal{B}^{(n)}_{(n)} =\mathcal{A}^{(n)}_{(n)}\colon  \mathcal{B}^{(n)} .
\end{equation} 
Inserting now the decomposition (\ref{harmonic decomposition tensor}) of $\mathcal{B}^{(n)}_{(n)}$ into the above result, and using once more the property (\ref{dot colon}) of the tensor contraction, we get
\begin{eqnarray} \label{harmonic contraction}
\mathcal{A}^{(n)}_{(n)}\colon  \mathcal{B}^{(n)}_{(n)} &=& \sum_{p =0}^{\lfloor n/2 \rfloor} 
\frac{p_{n,p}}{p_{n,0}} \, \mathcal{A}^{(n,p)}\colon  \mathcal{B}^{(n,p)} \\
&=&  \sum_{p =0}^{\lfloor n/2 \rfloor}
\frac{(-1)^p \, n! \, (2n-2p-1)!!}{2^p\, p!\, (2n-1)!! \, (n-2p)!}\, \mathcal{A}^{(n,p)}\colon  \mathcal{B}^{(n,p)}  .
\end{eqnarray} 
We have thus been able to express the contraction of the order-$n$ harmonic components of two rank-$n$ tensors explicitly, in terms of the contractions of their traces.

\section{Maxwell-Sylvester Multipoles}\label{sec_maxwell_sylvester}

\subsection{Heuristic Presentation of the Maxwell-Sylvester Multipoles}

So far, we have discussed spherical harmonics using only their general properties without specifying them. The traditional way of using them consists in expanding them on the basis of the familiar $2\ell +1$ complex $Y_\ell^m$ or real $Y_{\ell m}$ spherical harmonics \cite{cohen1977,jackson1999}. In this approach, a order-$\ell$ spherical harmonic is then given by the $2\ell +1$ complex expansions coefficients. This cartesian-algebraic approach has the advantages and inconvenients discussed in the Introduction.

An alternative approach based upon coordinate-free geometric concepts has been introduced by Maxwell some 125 years ago \cite{maxwell1891}. Maxwell's approach is most easily understood by considering the potential of electrostatic multipoles. As pointed out earlier (see Eq.~(\ref{potential harmonic})), each order-$\ell$ spherical harmonic is associated to the electrostatic potential of an order-$\ell$ multipole (a $2^\ell$-pole). Maxwell's construction can be physically understood in the following way: a dipolar electrostatic potential may be obtained by taking 2 opposite charges and shifting them by a small amount $\mathbf{u}_1$. Similarly, a quadrupolar field may be obtained by taking 2 opposite dipole moments and by shifting them by a small amount $\mathbf{u}_2$. The process can be iterated further and a $2^{\ell}$-polar potential may be obtained by shifting 2 opposite $2^{\ell -1}$-poles and shifting them by a small amount $\mathbf{u}_{\ell}$. Without any restriction, the vectors $\mathbf{u}_1$, $\mathbf{u}_2$, ..., $\mathbf{u}_\ell$ may be taken to be unit vectors by rescaling the charge $q$. So, Maxwell expresses an order-$\ell$ multipolar field as
\begin{equation}
V^{(\ell )}_{\mathsf{U}^{(\ell )}}(\mathbf{r}) = q \frac{(-1)^\ell}{\ell !} \prod_{i=1}^\ell \left( \hat{\mathbf{u}}_i \cdot \partial_\mathbf{r} \right) \frac{1}{r} ,
\end{equation}
where $\mathsf{U}^{(\ell )} \equiv \left\{ \hat{\mathbf{u}}_1, \hat{\mathbf{u}}_2 , \dots , \hat{\mathbf{u}}_\ell \right\}$ is an unordered collection (multiset) of (non-necessarily distinct) unit vectors. That the above expression is actually a harmonic function follows immediately from the fact $1/r$ is harmonic.
The importance of Maxwell's geometric multipole representation stems from a theorem due to Sylvester \cite{sylvester1876} who proved that any $2^\ell$-polar potential (i.e., any real order-$\ell$ solid harmonic) can be expressed by Maxwell's representation, which is actually unique (up to a change of sign of any pair of the unit vectors $\hat{\mathbf{u}}_i$).

\subsection{Maxwell-Sylvester Multipoles as a Tool for the Geometric Representation of Spherical Harmonics}
\label{sec_skeleton}

Let us now discuss the geometric interpretation of the Maxwell-Sylvester representation. It associates to each order-$\ell$ solid harmonic (or harmonic polynomial, or traceless symmetric tensor) with a multiset $\mathsf{U}^{(\ell )} \equiv \left\{\mathbf{u}_1, \dots , \mathbf{u}_\ell \right\}$. From the latter, one can construct the symmetric tensor
\begin{equation}
\mathcal{U}^{(\ell)}\equiv \bigodot_{i=1}^{\ell} \mathbf{u}_i ,
\end{equation}
which may vie as a "skeleton" of the associated harmonic tensor $\mathcal{U}^{(\ell)}_{(\ell)}$. The latter can be constructed by the method described in Sec.~\ref{section canonical} above. While the solid harmonic $\mathcal{U}^{(\ell)}_{\ell}(\hat{\mathbf{n}})$ may be a quite complicated function of $\hat{\mathbf{n}}$, its skeleton 
\begin{equation}
\mathcal{U}^{(\ell)}(\hat{\mathbf{n}}) = \prod_{i=1}^{\ell} (\mathbf{u}_i\cdot \hat{\mathbf{n}})
\end{equation}
is very simple as it vanishes on $\ell$ grand circles, which are the equators corresponding to the poles $\mathbf{u}_i$. So, if we represent the function $\mathrm{sgn}(\mathcal{U}^{(\ell)}(\hat{\mathbf{n}}))$ on a sphere, we obtain a very simple geometric representation which entirely determines the solid harmonic, up to a the scale factor
\begin{equation}
|\mathcal{U}^{(\ell)}| \equiv  \prod_{i=1}^\ell u_i ,
\end{equation}
which we may represent geometrically as the radius of the sphere. We note that, in the case of multiply degenerate equators, we would have to specify their multiplicity as well. 

At this point, it may be useful to make the link with the familiar (real) spherical harmonics $Y_{\ell, m}$, and specify how the latter are described in terms of the Maxwell-Sylvester representation. The zonal spherical harmonics $Y_{\ell, 0}$ have all their $\ell$ great circles degenerate and located on the "geographic" equator. The sectorial spherical harmonics $Y_{\ell ,\pm \ell}$ have their great circles located on "meridian lines" with an angle separation of $2\pi /\ell$ (the great circles for $m=-\ell$ are shifted by an angle $\pi\ell$ from those for $m=-\ell$). Finally, for a general value of $m$, there are $|m|$ great circle in meridian planes, separated by an angle $2\pi /m$, and $\ell - |m|$ degenerate great circles in the equatorial plane. 

The complex sectorial spherical harmonics $H_{\ell}^{\pm \ell}$ are particularly simple. They are given (up to the prefactor) by 
\begin{equation}
H_{\ell}^{\pm \ell} = (\hat{\mathbf{x}}\pm\mathrm{i}\hat{\mathbf{y}})^{\bigodot \ell}
\end{equation}
and are actually equal to their skeleton, because 
\begin{equation}
(\hat{\mathbf{x}}\pm\mathrm{i}\hat{\mathbf{y}})\cdot (\hat{\mathbf{x}}\pm\mathrm{i}\hat{\mathbf{y}}) =0 ,
\end{equation}
so that all the traces in the expansion (\ref{harmonic decomposition tensor}). 
Similarly, any tensor of the form 
\begin{equation}
(\hat{\mathbf{a}}\pm\mathrm{i}\hat{\mathbf{b}})^{\bigodot \ell} 
\end{equation}
with $a=b$ and $\hat{\mathbf{a}}\cdot \hat{\mathbf{b}}=0$ is a sectorial harmonic of axis $\hat{\mathbf{a}}\times \hat{\mathbf{b}}$.

Since the solid harmonic has the same symmetries as its skeleton, this representation is particularly well suited for analyzing its symmetries, unlike the traditional description of a real spherical harmonic in terms of its expansion over the basis of the (real) $H_l^m$. Conversely, constructing a spherical harmonic that possesses some particular symmetry is immediate and reduces to a geometric construction on the sphere, whereas except in some simple cases, the construction of spherical harmonics can be a tedious exercise, and the symmetry of the obtained result is frequently not manifestly apparent. Furthermore, the Maxwell-Sylvester geometric representation offers a very simple and natural way of constructing solid harmonics that deviates slightly from a given symmetry, by moving slightly the multipole vectors away from their symmetric positions. 

\subsection{Electrostatic Interpretation of the Maxwell-Sylvester Multipole Vectors}

There is a very interesting electrostatic analogy associated with the Maxwell-Sylvester representation. Let us consider a sphere with a surface charge distribution $\sigma^{(\ell)}_\mathcal{A} (\hat{\mathbf{n}})$ given by the spherical harmonic $\mathcal{A}^{(\ell)}_{(\ell)}$. From Poisson's equation, it follows that the electrostatic energy is (up to some, for us unimportant, factor)
\begin{eqnarray}
E_\mathcal{A} &=& \frac{1}{2 \ell (\ell +1)} \, \frac{1}{4\pi} \int_{S^2}\!\!\!\mathrm{d}^2 \hat{\mathbf{n}}\ |\sigma^{(\ell)}_\mathcal{A} (\hat{\mathbf{n}})|^2  \\
&=& \frac{1}{2 \ell (\ell +1)} \, \left\langle  \mathcal{A}^{(\ell)}_{(\ell)} | \mathcal{A}^{(\ell)}_{(\ell)} \right\rangle .
\end{eqnarray}
so the electrostatic energy of a multipole is expressed by its norm. So, if we build the superposition of two such multipoles $\mathcal{A}^{(\ell)}_{(\ell)}$ and $\mathcal{B}^{(\ell)}_{(\ell)}$, the energy of the resultant multipole is
\begin{eqnarray}
E_{\mathcal{A}+\mathcal{B}} &=& \frac{1}{2 \ell (\ell +1)} \, \frac{1}{4\pi} \int_{S^2}\!\!\!\mathrm{d}^2 \hat{\mathbf{n}}\ |\sigma^{(\ell)}_\mathcal{A} (\hat{\mathbf{n}}) +|\sigma^{(\ell)}_\mathcal{B} (\hat{\mathbf{n}}) |^2  \\
&=& \frac{1}{2 \ell (\ell +1)} \, \left(  \left\langle  \mathcal{A}^{(\ell)}_{(\ell)} | \mathcal{A}^{(\ell)}_{(\ell)} \right\rangle  + \left\langle  \mathcal{B}^{(\ell)}_{(\ell)} | \mathcal{B}^{(\ell)}_{(\ell)} \right\rangle + 2 \left\langle  \mathcal{A}^{(\ell)}_{(\ell)} | \mathcal{B}^{(\ell)}_{(\ell)} \right\rangle \right) .
\end{eqnarray}
So, the mechanical work $\Delta E_{\mathcal{A},\mathcal{B}}$ needed to superpose the two multipoles by bringing one onto another from infinite distance (i.e., their mutual interaction energy) is expressed by the inner product of the two multipoles. 

Let us consider a real sectorial harmonic
\begin{equation}
\mathcal{C}^{(\ell)}_{(\ell)} = \mathrm{Re}\left[  \left( (\hat{\mathbf{a}}+\mathrm{i}\hat{\mathbf{b}})\mathrm{e}^{\mathrm{i}\phi}\right) ^{\odot \ell}\right]  ,
\end{equation}
with $a=b$ and $\hat{\mathbf{a}}\cdot \hat{\mathbf{b}}=0$. As discussed earlier, it is a sectorial multipole of axis $\hat{\mathbf{a}}\times \hat{\mathbf{b}}$, with an azimuthal angle determined by the phase $\phi$. 

Let us consider the interaction energy between a generic multipole $\mathcal{U}^{(\ell)}_{(\ell)}$ given as the traceless component of a symmetrized tensor product
\begin{equation}
\mathcal{U}^{(\ell)}_{(\ell)} \equiv \bigodot_{i=1}^{\ell} \mathbf{u}_i
\end{equation}
and the above sectorial multipole. Using the formulae Eq.~(\ref{inner product harmonics},\ref{harmonic contraction}) and because all the traces of the sectorial multipole vanish, we have
\begin{equation}
\Delta E_{\mathcal{U},\mathcal{C}} = \frac{1}{2 \ell (\ell +1)} \, 
\frac{\ell!}{(2\ell +1)!!} \mathrm{Re}\left[    \mathrm{e}^{\mathrm{i}\ell\phi}
 \prod_{i=1}^{\ell} (\mathbf{u}_i \cdot (\hat{\mathbf{a}}+\mathrm{i}\hat{\mathbf{b}}))
\right] .
\end{equation}
So, if the axis $\hat{\mathbf{a}}\times \hat{\mathbf{b}}$ of the sectorial multipole is parallel to any of the of the multipole vectors $\mathbf{u}_i$, independently of the azimuthal angle $\phi$, the interaction between the multipole $\mathcal{U}^{(\ell)}_{(\ell)}$ and the sectorial multipole $\mathcal{C}^{(\ell)}_{(\ell)}$ vanishes. Thus, from Sylvester's theorem we see that for any real $2^\ell$-pole, there are exactly $\ell$ (possibly) degenerate sectorial multipoles axes which have vanishing interaction energy with $\mathcal{U}^{(\ell)}_{(\ell)}$. This interpretation shows that the Maxwell-Sylvester multipole vector have a clear physical reality and indicates (at least conceptually) how they can be determined.

Surprisingly, more than 150 years after its discovery by Maxwell and Sylvester, the geometric representation of multipoles has remained widely ignore by the physics community, and has found very little application in physics. The main use of the Maxwell-Sylvester multipoles, so far, has been for the symmetry classification of elasticity tensors \cite{Backus1970,zou2002}. 
Recently, it has found some renewed interest and been used to analyze and interpret the anisotropy of the cosmic microwave background radiation temperature \cite{copi2004,dennis2008}. 

However, it has so far found no use in the quantum theory of atoms, nuclei and molecules, and their interaction with electromagnetic radiation, in spite of the prominent role played by the concept of multipoles in this context. To provide the basis for a geometric theory of quantum multipoles constitutes the main objective of the present work.

\section{Geometric representation of quantum spin states}\label{sec_coherent_majorana}

Let us now consider a system consisting of a single quantum spin of magnitude $J$ (which may be integer or half-integer). We are interested only in the degrees of freedom associated with the angular momentum $J$, so that the Hilbert space is $\mathcal{H}\equiv\{\mathbb{C}^{(2J+1)} -0\}$. The traditional way of describing a (pure) quantum state of the systems consists in specifying its $(2J+1)$ complex components on some basis of the Hilbert space (most frequently, this is the familiar basis of the eigenstates $\left| J,m\right\rangle $ of the $J_z$ operator). Operators describing physical observables, on the other hand, are represented by $(2J+1)\times (2J+1)$ Hermitian matrices acting on the above vectors states. While it is convenient for calculations, such "cartesian" description of the quantum states usually conveys very little physical insight, except in the simplest cases. 

On a more fundamental level, a further weakness lies in the fact that since two Hilbert space vectors differing only by a (non-zero) complex prefactor lead to the same expectation value for any physical observable, they should be considered as actually representing the same physical state of the system; this means that the Hilbert space $\mathcal{H}$ is \emph{not} the actual space of physical states of the quantum system; instead the phase space of physical (pure) quantum states is the projective Hilbert space $\mathcal{PH} = \mathcal{H}/\sim$, the quotient set of equivalence classes in $\mathcal{H}$, where $\left| \psi_1 \right> \sim \left| \psi_2 \right>$ iff $\left| \psi_1 \right> = c \left| \psi_2 \right>$ (with $c\in \mathbb{C}, c\neq0$). For example, for a Hilbert space of finite dimension $(2J+1)$, the projective Hilbert space is $\mathcal{PH}=\mathbb{C}P^{2J}$, the complex projective space of complex dimension $2J$. Because of this redundancy of the Hilbert space, using it to formulate quantum mechanics does not allow to reveal in full depth the mathematical structure of quantum mechanics.

It is therefore desirable to have a geometric description of the quantum states of the system, that would be independent of a particular choice of coordinates and basis states, and that would provide a much more natural way of investigating its symmetries, while still enabling to perform conveniently all physically relevant calculations. 

\subsection{Spin Coherent States}

One such geometric description is that of the spin coherent states $\left| \hat{\mathbf{n}}^{(J)} \right\rangle $ \cite{arecchi1972,gilmore1975,perelomov1986}, which are obtained by rotating the fully polarized state $\left| \hat{\mathbf{z}}^{(J)} \right\rangle \equiv \left| J,J\right\rangle $ from the $\hat{\mathbf{z}}$ axis to the $\hat{\mathbf{n}}$ axis, i.e.,  $\left| \hat{\mathbf{n}}^{(J)} \right\rangle $ is defined (up to a norm and phase convention) by
\begin{equation}
\left( \mathbf{J}\cdot \hat{\mathbf{n}} \right) 
\left| \hat{\mathbf{n}}^{(J)} \right\rangle = J\, \left| \hat{\mathbf{n}}^{(J)} \right\rangle .
\end{equation}
Their scalar product is given by
\begin{equation}
\left\langle \hat{\mathbf{n}}^{(J)} | \hat{\mathbf{n}}^{\prime(J)} \right\rangle =
\left( \frac{1+\hat{\mathbf{n}}\cdot \hat{\mathbf{n}}^{\prime}  }{2}\right)  \, \mathrm{e}^{\mathrm{i} J \Sigma (\hat{\mathbf{z}},\hat{\mathbf{n}},\hat{\mathbf{n}}^{\prime})} ,
\end{equation}
where $\Sigma (\hat{\mathbf{z}},\hat{\mathbf{n}},\hat{\mathbf{n}}^{\prime})$ is the oriented area of the spherical triangle  $(\hat{\mathbf{z}},\hat{\mathbf{n}},\hat{\mathbf{n}}^{\prime})$. The above expression fixes the norm and phase convention. The spin coherent states satisfy the following resolution of unity
\begin{equation}
\mathbf{1}_{J} \equiv \frac{2J+1}{4\pi}\int_{S^2} \mathrm{d}^{2}\mathbf{\hat{n}} \ |\mathbf{\hat{n}}^{(\! J)}\rangle \langle \mathbf{\hat{n}}^{(\! J)}| ,
\end{equation}
which enables to represent a pure quantum state $\left| \Psi^{(J)}\right\rangle $ by its projection on the coherent states $\Psi^{(J)}(\mathbf{\hat{n}})\equiv \langle \mathbf{\hat{n}}^{( J)}| \Psi^{(\! J)}\rangle$, which is a wave function over the sphere $S^2$. Actually, because of the redundancy of the continuous coherent states (by contrast with the projection on the (2J+1) states of an orthonormal basis), the phase information is not needed and all observables can be obtained from from the Husimi function
\begin{equation}
Q_{\Psi^{(J)}} (\mathbf{\hat{n}}) \equiv  |\Psi^{(\! J)}(\mathbf{\hat{n}})|^{2} ,
\end{equation}
which is a probability distribution on the sphere. Clearly, the coherent states representation has the desired geometric character. In particular all the symmetry properties of the state $\left| \Psi^{(J)}\right\rangle $ will be manifestly displayed as symmetries of its representation on the sphere via $Q_{\Psi^{(J)}} (\mathbf{\hat{n}}) $.

The coherent states representation has found considerable use and is a central tool in the theory of quantum spin systems \cite{auerbach1994,nagaosa1999}. In particular it is at the heart of the path integral theory for quantum spin systems.

\subsection{Majorana's Stellar Representation}

On the other hand, except for spin-$1/2$ systems, the coherent states constitutes a small sub-manifold  of the projective Hilbert space, and, at first sight, it might seem that the detailed description of the Husimi distribution on the sphere would be needed; this is of course not true, and the Husimi distribution highly redundant.  It is therefore desirable to have a more compact geometric description of a quantum state. This is precisely what is achieved by Majorana's stellar representation \cite{majorana1932,bloch1945,schwinger1952}. The latter relies on a theorem due to Majorana's which establishes that the wavefunction $\Psi^{(J)}(\mathbf{\hat{n}})$ has exactly $2J$ (possibly degenerate) zeros, and that these zeros entirely determine the quantum state represented by   $| \Psi^{(\! J)}\rangle$. So the state $| \Psi^{(\! J)}\rangle$ can be represented geometrically by a "constellation" of $2J$ "stars" (the antipodal points of the zeros of $\Psi^{(J)}(\mathbf{\hat{n}})$) on the sphere. The Majorana representation may be thought of as the generalization for a spin $J >1/2$ of Bloch's representation of a spin-$1/2$ system, in which a spin-$1/2$ state is described geometrically as a coherent state $\left| \hat{\mathbf{n}}^{(1/2)} \right\rangle $. Since a spin-$J$ system may be constructed as a fully symmetrized (i.e., bosonic) state of a composite system made of $2J$ spins-$1/2$, the Majorana representation of the spin-$J$ systems is the constellation of the $2J$ Bloch vectors of its $2J$ spin-$1/2$ constituents. The striking similarity between the geometric representations of Majorana for pure quantum spin states and of Maxwell-Sylvester for spherical harmonics is no coincidence, and reflects the fact that for $J$ integer, the time-reversal-invariant quantum states are given by real spherical harmonics, so that Majorana's problem reduces to the Maxwell-Sylvester problem. In this case, the constellation of the $2J$ Majorana stars constitutes in a multiset of $J$ pairs of antipodal stars, whose great circles (to which they are perpendicular) coincide with the Maxwell-Sylvester representation \cite{dennis2004}. The Majorana representation has been used in particular to shed light on the geometric phase and dynamics of quantum spins \cite{bruno2012} and to discuss the "anti-coherence" of symmetric multi-qubits systems \cite{giraud2015,baguette2015,baguette2015}. We shall not discuss it further here, and use essentially the coherent state representation.

\section{Geometric Representation of the Irreducible Tensor Operators}\label{sec_geometric_operators}

Besides the description of the quantum states themselves, we also need a geometric description of the operators acting in the Hilbert space and representing physical observable quantities. The traditional approach to this relies on Stevens' construction of irreducible tensor operators \cite{stevens1952,buckmaster1962,buckmaster1971,buckmaster1972,rudowicz2004}. In this approach, physical observables (such as the Hamiltonian) are constructed by quantizing a classical function of the angular momentum: $A_\text{cl}(\mathbf{j})$. The latter can be expressed as a polynomial of degree $n$ of the components $j_x,j_y,j_z$ of the classical angular momentum, which can be expanded in harmonic homogeneous polynomials:
\begin{equation}
A_\text{cl}(\mathbf{j}) = \sum_{\ell=0}^{n}A_{\text{cl}(\ell)}(\mathbf{j}) .
\end{equation}
where $A_{\text{cl}(\ell)}$ is an order-$\ell$ solid harmonic, which is usually expanded on the basis of the familiar (real) solid harmonics $H_{\ell,m}$.  
The corresponding quantum operator is obtained by substituting the classical angular momentum by the quantum mechanical operator $\hat{\mathbf{J}}$, i.e.,
\begin{equation}
\hat{\mathcal{A}}(\mathbf{J}) = \sum_{\ell=0}^{n}\hat{\mathcal{A}}_{(\ell)}(\hat{\mathbf{J}}) .
\end{equation}
In the above expression, a symmetrization of all products of the angular momentum components is meant. As the various components of $\hat{\mathbf{J}}$ do not commute, the symmetrization ensures that (i) the resulting operator is Hermitian as required for a physical observable, and (ii) that the degree of the polynomial cannot be lowered by use of the commutation relations. Thus, one has to carry out the task of performing explicitly this symmetrization for the solid harmonics 
$H_{\ell,m}$, in order to generate the Stevens operators. This is fairly easy for $\ell \le 3$ but rapidly gets highly complicated for the higher values of $\ell$ (see e.g. \cite{rudowicz2004}). 

The geometric approach relies on a description in terms of the coherent states, instead of a fixed basis as in the "cartesian" approach. To the operator $\hat{\mathcal{A}}(\mathbf{J})$ we associate the spherical function
\begin{equation}
a(\hat{\mathbf{n}})  \equiv \left\langle J,\hat{\mathbf{n}}\right|  \hat{\mathcal{A}}(\mathbf{J}) \left| J,\hat{\mathbf{n}}\right\rangle ,
\end{equation}
which is called the Q-representation of the operator $\hat{\mathcal{A}}(\mathbf{J})$ \cite{gilmore1976}. There is also a (non-unique) P-representation $A(\hat{\mathbf{n}})$ defined by
\begin{equation}
\hat{\mathcal{A}}(\mathbf{J}) \equiv \frac{(2J+1)}{4\pi}\int_{S^2}\!\!\!\!\mathrm{d}^2\hat{\mathbf{n}}\ 
\left| J,\hat{\mathbf{n}}\right\rangle A(\hat{\mathbf{n}}) \left\langle J,\hat{\mathbf{n}}\right| .
\end{equation}
The non-unicity of the P-representation can be fixed by requiring 
\begin{equation}
\int_{S^2}\!\!\!\!\mathrm{d}^2\hat{\mathbf{n}}\ A(\hat{\mathbf{n}}) P_\ell (\hat{\mathbf{n}} \cdot \hat{\mathbf{u}}) =0, \ \ \ \forall \hat{\mathbf{u}} \in \, S^2, \ \text{if} \ \ell > 2J .
\end{equation}
Unlike the representation of the observable in terms of its matrix elements in a fixed basis, in which all matrix elements (on-diagonal and off-diagonal) are needed to fully specify the observable, in the coherent state representation, due to the redundancy of the coherent states, the observable is fully specified by its diagonal elements only.
 
let us expand $a(\hat{\mathbf{n}})$, and $A(\hat{\mathbf{n}})$ into their harmonic components:
\begin{eqnarray}
a(\hat{\mathbf{n}})&= & \sum_{\ell} a_\ell(\hat{\mathbf{n}})  \\
A(\hat{\mathbf{n}})&= & \sum_{\ell} A_\ell(\hat{\mathbf{n}}) ,
\end{eqnarray}
where
\begin{eqnarray}
a_\ell(\hat{\mathbf{n}}) &=& \frac{2\ell +1}{4\pi}\!\!\int_{S^2}\!\!\!\!\mathrm{d}^2\hat{\mathbf{u}}\ a(\hat{\mathbf{u}}) P_\ell (\hat{\mathbf{u}}\cdot \hat{\mathbf{n}}) \\
A_\ell(\hat{\mathbf{n}}) &=& \frac{2\ell +1}{4\pi}\!\!\int_{S^2}\!\!\!\!\mathrm{d}^2\hat{\mathbf{u}}\ A(\hat{\mathbf{u}}) P_\ell (\hat{\mathbf{u}}\cdot \hat{\mathbf{n}}) .
\end{eqnarray}
The problem is now to related the $a_\ell(\hat{\mathbf{n}})$ and $A_\ell(\hat{\mathbf{n}})$ to the harmonic components of the classical potential  $A_{\text{cl}\ell}(\hat{\mathbf{n}})$ of the classical potential. From considerations of their rotational symmetry behavior, it follows that must must be proportional to each other, i.e.,
\begin{eqnarray}
a_\ell(\hat{\mathbf{n}}) &=& \alpha_{J,\ell} \, A_{\text{cl}\ell}(\hat{\mathbf{n}}) \\
A_\ell(\hat{\mathbf{n}}) &=& \beta_{J,\ell} A_{\text{cl}\ell}(\hat{\mathbf{n}}) ,
\end{eqnarray}
where the coefficients $\alpha{J,\ell}$ and $\beta{J,\ell}$ are given by \cite{gilmore1976}.
\begin{eqnarray}
\alpha_{J,\ell} &=& \frac{(2J)!}{2^\ell \, (2J-\ell)!} \ \ \text{for}\ \ \ell \le 2J, \\
\beta_{J,\ell} &=& \frac{(2J+\ell +1)!}{2^\ell \, (2J+1)!} \ \ \text{for}\ \ \ell \le 2J, \\
\alpha_{J,\ell} &=& \beta_{J,\ell} =0\ \ \text{for}\ \ \ell > 2J.
\end{eqnarray}

If we now use the Maxwell-Sylvester geometric representation described in Sec.~\ref{sec_skeleton} $A_\ell(\hat{\mathbf{n}})$ (or equivalently, for $a_\ell(\hat{\mathbf{n}})$, or $A_{\text{cl}\ell}(\hat{\mathbf{n}})$), if we represent them in terms of the skeleton on the sphere, we now describe any operator by a collection of geometrical patterns consisting of sets of great circles on spheres, that contain all the information we may need concerning our operator. At the same time, this representation displays directly all the symmetries of the operator.

Using this, we can now express the expectation value of the operator for a quantum state $\left| \Psi^{(J)} \right\rangle$ (which we take normalized to 1)
\begin{eqnarray}
\left\langle \Psi^{(J)} \right| \hat{\mathcal{A}}(\mathbf{J}) \left| \Psi^{(J)} \right\rangle &=& 
\frac{(2J+1)}{4\pi}\int_{S^2}\!\!\!\!\mathrm{d}^2\hat{\mathbf{n}}\ 
\left\langle \Psi^{(J)} \right. \left| J,\hat{\mathbf{n}}\right\rangle A(\hat{\mathbf{n}}) \left\langle J,\hat{\mathbf{n}}\right. \left| \Psi^{(J)} \right\rangle \\
&=& \frac{(2J+1)}{4\pi}\int_{S^2}\!\!\!\!\mathrm{d}^2\hat{\mathbf{n}}\ 
Q^{(\! J)}_\Psi (\mathbf{\hat{n}})\, A(\hat{\mathbf{n}}) \\
&=& \sum_{\ell =0}^{2J} (2J+1)\, \frac{1}{4\pi}\int_{S^2}\!\!\!\!\mathrm{d}^2\hat{\mathbf{n}}\ 
Q_{\Psi^{(J)}  (\ell)}  (\mathbf{\hat{n}})\, A_{(\ell)}(\hat{\mathbf{n}}) . 
\end{eqnarray}
We now use the Maxwell-Sylvester representation for both $Q_{\Psi^{(J)}  (\ell)}$ and $A_{(\ell)}$ and can apply the formulae (\ref{inner product harmonics},\ref{harmonic contraction}). The skeletons of $A_{(\ell)}$ and $Q^{(\! J)}_{\Psi (\ell)}$ are, respectively
\begin{eqnarray}
A^{(\ell)} &\equiv & \bigodot_{i=1}^{\ell}\ \mathbf{a}_i^{(\ell)}  \\
Q_{\Psi^{(J)}}^{(\ell)} &\equiv & \bigodot_{i=1}^{\ell}\ \mathbf{q}_i^{(\ell)}.
\end{eqnarray}

The final expression for the expectation value of the operator is
\begin{equation}
\left\langle \Psi^{(J)} \right| \hat{\mathcal{A}}(\mathbf{J}) \left| \Psi^{(J)} \right\rangle = 
(2J+1)  \sum_{\ell = 0}^{2J}\frac{\ell!}{(2\ell +1)!!} \sum_{p =0}^{\lfloor \ell/2 \rfloor} 
\frac{p_{\ell,p}}{p_{\ell,0}} \,Q_{\Psi^{(J)}}^{(\ell ,p)}
\colon A^{(\ell ,p)} ,
\end{equation}
where the traces are given by
\begin{equation}
A^{(\ell ,p)} = \frac{1}{\ell!}\sum_{\sigma \in S_\ell}\, \left( \prod_{i=1}^{p} (\mathbf{a}^{(\ell)}_{\sigma (i)} \cdot \mathbf{a}^{(\ell)}_{\sigma (i+p)})\right)  \, \left( \bigodot_{j=2p+1}^{\ell}\mathbf{a}^{(\ell)}_{\sigma (j)}\right)  ,
\end{equation}
with analogous expression for $Q_{\Psi^{(J)}}^{(\ell ,p)}$. The contraction of the traces is 
\begin{eqnarray}
Q_{\Psi^{(J)}}^{(\ell ,p)}
\colon A^{(\ell ,p)} &=& \frac{1}{(\ell!)^2}\sum_{\sigma, \sigma^\prime \in S_\ell} \left( \prod_{i=1}^{p} (\mathbf{q}^{(\ell)}_{\sigma (i)} \cdot \mathbf{q}^{(\ell)}_{\sigma (i+p)})\right)
\left( \prod_{j=1}^{p} (\mathbf{a^{(\ell)}}_{\sigma^\prime (j)} \cdot \mathbf{a}^{(\ell)}_{\sigma^\prime (j+p)})\right) \times \\
&&\times \left( 
\prod_{k=2p+1}^{\ell} (\mathbf{q}^{(\ell)}_{\sigma (k)} \cdot \mathbf{a}^{(\ell)}_{\sigma^\prime (k)})
\right) ,
\end{eqnarray}
which completes the announced result.

\section*{Appendix}

\subsection{Notations and conventions}

Vectors are noted using boldface l.c. letters ($\mathbf{a},\mathbf{b} $, etc.). Unit vectors, i.e., vectors belonging to the 2-sphere of unit radius $S^2$, will be indicated by a "hat" $\hat{\mathbf{n}}$. 
Capital boldface letters are used for ordered collections of vectors, e.g. $\mathbf{V}^{(n)}\equiv \{\mathbf{v}_1,\mathbf{v}_2, \dots , \mathbf{v}_n \}$. Caligraphic capitals $\mathcal{A}^{(n)}$ are used for tensors, the rank of which is indicated by the upper index $(n)$. We use the symbol $\otimes$ to represent the tensor product of two tensors, e.g.  $\mathcal{A}^{(n)}\otimes \mathcal{B}^{(m)}$, and the symbol $\odot$ to represent the fully symmetrized tensor product, as defined explicitly in Eq.~(\ref{sym product}).

\subsection{Legendre Polynomials}

As Legendre polynomials play an important role in this paper, some of their properties are reminded here \cite{Gradshteyn2007}. The Legendre polynomial of order $\ell$ can be defined by Rodrigues'formula \cite{GR-8.910-2}
\begin{equation}
P_\ell (x) \equiv \frac{1}{2^\ell} \frac{\mathrm{d}^\ell}{\mathrm{d}x^\ell}\left( x^2 -1 \right)^\ell .
\end{equation}
Their explicit expression is \cite{GR-8.911-1}
\begin{equation} \label{legendre expansion}
P_\ell(x)\equiv \sum_{k=0}^{\lfloor \ell /2\rfloor}p_{\ell, k} x^{\ell-2k} ,
\end{equation}
where the floor function is defined by 
\begin{equation}
\lfloor x \rfloor \equiv n, \text{ for } n\le x < n+1, \ (n\in \mathbb{Z})
\end{equation}
and where
\begin{equation}
p_{\ell, k} \equiv \frac{(-1)^k (2\ell-2k-1)!!}{2^k k! (\ell -2k)!} .
\end{equation}
Legendre polynomial or even (odd) order contain only monomials of even (odd) order. They are orthogonal to each other \cite{GR-7.112-1}:
\begin{equation}\label{legendre orthogonality}
\frac{1}{2} \int_{-1}^{1} \!\!\mathrm{d}x\ P_\ell (x) P_{\ell^\prime} (x) = \frac{\delta_{\ell \ell^\prime}}{2\ell +1} .
\end{equation}
From the latter relation, we also obtain the completeness relation
\begin{equation} \label{completeness_Legendre}
\delta (x-y) = \sum_{\ell = 0}^{\infty} \left( \ell + \frac{1}{2}\right)  P_\ell (x) P_\ell (y) .
\end{equation}
We shall also need the reciprocal of Eq.~(\ref{legendre expansion}), the expression of monomials $x^n$ in terms of Legendre polynomials. Using Eqs.(\ref{legendre orthogonality},\ref{completeness_Legendre}) and taking into account the parity of the Legendre polynomials, we obtain
\begin{equation}\label{monomial}
x^n = \sum_{k=0}^{\lfloor n /2\rfloor}q_{n, k} P_{n-2k}(x),
\end{equation}
with
\begin{equation}
q_{n, k} \equiv \frac{2(n-2k)+1}{2}\int_{-1}^{1} \mathrm{d}x\ x^n P_{n-2k}(x) .
\end{equation}
The calculating explicitly the integral \cite{GR-7.126-1}, we finally obtain
\begin{equation}\label{qnk}
q_{n, k} = \frac{(2n-4k+1)\, n!}{2^k\, k!\, (2n-2k+1)!!  } .
\end{equation}
Note that 
\begin{eqnarray}
p_{n,0}=\dfrac{1}{q_{n,0}}= \frac{(2n-1)!!}{n!} .
\end{eqnarray}

We shall also need the expansion of $(1+x)^n$ in Legendre polynomials. Proceeding like above, and using the formula \cite{GR-7.127}
\begin{equation}
\int_{-1}^{1}dx\ (1+x)^n\, P_\ell(x) = \frac{2^{n+1} (n!)^2}{(n+\ell+1)!\, (n-\ell)!} ,
\end{equation}
we obtain
\begin{equation}
(1+x)^n = 2^n \,(n!)^2 \sum_{\ell=0}^{n} \frac{2l+1}{(n+\ell +1)!\, (n-\ell)!} \, P_\ell(x) .
\end{equation}

Using the generating function \cite{GR-8.921}
\begin{equation}
\frac{1}{\sqrt{1-2t x+t^2}}=\sum_{\ell = 0}^{\infty} t^\ell\ P_\ell (x) ,
\end{equation}
one can show easily that
\begin{eqnarray}
P_\ell\left(\frac{\mathbf{r}\cdot\mathbf{r}^\prime}{r r^\prime}\right)&=&\frac{(-1)^\ell}{\ell !} \frac{r^{\prime\,\ell+1}}{r^\ell} \left(\mathbf{r}\cdot\frac{\partial}{\partial\mathbf{r}^\prime}\right)^\ell \left(\frac{1}{r^\prime}\right) \\
&=&\frac{(-1)^\ell}{\ell !} \frac{r^{\ell+1}}{r^{\prime\,\ell}} \left(\mathbf{r}^\prime \cdot\frac{\partial}{\partial\mathbf{r}}\right)^\ell \left(\frac{1}{r}\right) ,
\end{eqnarray}
where $\mathbf{r}$ and $\mathbf{r}^\prime$ are vectors in $\mathbb{R}^3$. Since the electrostatic potential of a point charge satisfies Laplace's equation (except at the charge's location), i.e,
\begin{equation}
\Delta_\mathbf{r}\left( \frac{1}{r}\right)  =0 ,
\end{equation}
it follows immediately that the function
\begin{equation}
Y_\ell (\mathbf{r},\mathbf{r}^\prime)\equiv P_\ell \left( \frac{\mathbf{r}\cdot\mathbf{r}^\prime}{rr^\prime} \right) 
\end{equation}
is a spherical harmonic of order $\ell$ of $\mathbf{r}$ and $\mathbf{r}^\prime$, and 
\begin{equation}
H_\ell (\mathbf{r},\mathbf{r}^\prime)\equiv  r^\ell r^{\prime\ell} Y_\ell (\mathbf{r},\mathbf{r}^\prime)
\end{equation}
is a solid harmonic of order $\ell$ of $\mathbf{r}$ and $\mathbf{r}^\prime$, i.e.,
\begin{eqnarray}
&& \Delta_{\mathbf{r}} Y_\ell (\mathbf{r},\mathbf{r}^\prime) + \frac{\ell (\ell +1)}{r^2} Y_\ell (\mathbf{r},\mathbf{r}^\prime)=  \Delta_{\mathbf{r}^\prime} Y_\ell (\mathbf{r},\mathbf{r}^\prime)+ \frac{\ell (\ell +1)}{r^{\prime 2}} Y_\ell (\mathbf{r},\mathbf{r}^\prime) = 0 \\
&& \Delta_{\mathbf{r}} H_\ell (\mathbf{r},\mathbf{r}^\prime)=  \Delta_{\mathbf{r}^\prime} H_\ell (\mathbf{r},\mathbf{r}^\prime) = 0 .
\end{eqnarray}

Let us consider the function $f_{\ell,\ell^\prime}(\hat{\mathbf{u}},\hat{\mathbf{v}})$ of $(\hat{\mathbf{u}},\hat{\mathbf{v}}) \in S^2$ defined by
\begin{equation}
f_{\ell,\ell^\prime}(\hat{\mathbf{u}},\hat{\mathbf{v}}) \equiv \int_{S^2}\!\!\!\!\mathrm{d}^2\hat{\mathbf{n}}\ 
P_\ell (\hat{\mathbf{u}}\cdot\hat{\mathbf{n}}) \ P_{\ell^\prime} (\hat{\mathbf{n}}\cdot\hat{\mathbf{v}}) .
\end{equation}
It is obviously invariant by rotation, i.e., for any rotation $R \in SO(3)$, we must have
\begin{equation}
f_{\ell,\ell^\prime}(R(\hat{\mathbf{u}}),R(\hat{\mathbf{v}})) = f_{\ell,\ell^\prime}(\mathbf{u},\mathbf{v}) .
\end{equation}
In particular, if we chose the rotation which permutes $\hat{\mathbf{u}}$ and $\hat{\mathbf{v}}$, we have
\begin{equation}
f_{\ell,\ell^\prime}(\hat{\mathbf{v}},\hat{\mathbf{u}}) = f_{\ell,\ell^\prime}(\hat{\mathbf{u}},\hat{\mathbf{v}}) = f_{\ell^\prime,\ell}(\hat{\mathbf{u}},\hat{\mathbf{v}}).
\end{equation}
By construction $f_{\ell,\ell^\prime}(\hat{\mathbf{u}},\hat{\mathbf{v}})$ is a spherical harmonic of order $\ell$ with respect to its first argument and a spherical harmonic of order $\ell^\prime$ with respect to its second argument. Thus, we have
\begin{equation}
\ell (\ell +1) f_{\ell,\ell^\prime}(\hat{\mathbf{u}},\hat{\mathbf{v}}) = \ell^\prime (\ell^\prime +1) f_{\ell,\ell^\prime}(\hat{\mathbf{u}},\hat{\mathbf{v}}) ,
\end{equation}
which implies that $f_{\ell,\ell^\prime}(\hat{\mathbf{u}},\hat{\mathbf{v}})=0$ if $\ell \neq \ell^\prime$. We have thus established the orthogonality of spherical harmonics of different orders, and in particular
\begin{equation}\label{reproducing kernel}
\frac{1}{4\pi}\int_{S^2}\!\!\!\!\mathrm{d}^2\hat{\mathbf{n}}\ 
P_\ell (\hat{\mathbf{u}}\cdot\hat{\mathbf{n}}) \ P_{\ell^\prime} (\hat{\mathbf{n}}\cdot\hat{\mathbf{v}}) = \frac{\delta_{\ell\ell^\prime}}{2\ell +1}P_\ell (\hat{\mathbf{u}}\cdot\hat{\mathbf{v}})
\end{equation}

For any function $f(\hat{\mathbf{n}})$ defined on $S^2$, where $S^2$ is the unit-radius sphere in $\mathbb{R}^3$, the function
\begin{equation}\label{spherical harmonic component}
f_\ell (\hat{\mathbf{n}}) \equiv \frac{2\ell +1}{4\pi}\!\!\int_{S^2}\!\!\!\!\mathrm{d}^2\hat{\mathbf{u}}\ f(\hat{\mathbf{u}}) P_\ell (\hat{\mathbf{u}}\cdot \hat{\mathbf{n}})
\end{equation}
is obviously a spherical harmonic of order $\ell$. Thus, the above convolution of a spherical function with the order-$\ell$ Legendre polynomial enables to extract its order-$\ell$ harmonic component. Thus, a function $Y_\ell (\hat{\mathbf{n}})$ is a spherical harmonic iff it satisfies
\begin{equation}
Y_\ell (\hat{\mathbf{n}}) \equiv \frac{2\ell +1}{4\pi}\!\!\int_{S^2}\!\!\!\!\mathrm{d}^2\hat{\mathbf{u}}\ Y_l(\hat{\mathbf{u}}) P_\ell (\hat{\mathbf{u}}\cdot \hat{\mathbf{n}}) .
\end{equation}
In particular, we have
\begin{equation}
Y_\ell (\hat{\mathbf{n}}) \equiv \frac{2\ell +1}{4\pi}\!\!\int_{S^2}\!\!\!\!\mathrm{d}^2\hat{\mathbf{u}}\ Y_l(\hat{\mathbf{u}}) P_\ell (\hat{\mathbf{u}}\cdot \hat{\mathbf{n}}) .
\end{equation}
So, the Legendre polynomial $P_\ell (\hat{\mathbf{u}}\cdot \hat{\mathbf{n}})$ is the reproducing kernel of the space of order-$\ell$ spherical harmonics. From the completeness relation (\ref{completeness_Legendre}), we obtain
\begin{equation}
\delta (\hat{\mathbf{u}} -\hat{\mathbf{v}}) = \frac{1}{4\pi}\sum_{\ell =0}^{\infty} (2\ell +1) P_\ell (\hat{\mathbf{u}}\cdot \hat{\mathbf{v}}) .
\end{equation}
So any function on $S^2$ may be decomposed in a unique way into its harmonic components
\begin{equation}\label{spherical harmonic decomposition}
f(\hat{\mathbf{n}}) = \sum_{\ell = 0}^{\infty} f_\ell (\hat{\mathbf{n}}) .
\end{equation}
%


\end{widetext}

\end{document}